\documentclass[journal=jpccck,manuscript=article]{achemso}

\usepackage[version=3]{mhchem} % Formula subscripts using \ce{}
\usepackage{mathtools}
\usepackage{amssymb}
\usepackage{amsthm}
\usepackage{amsfonts}
\usepackage{braket}
\usepackage{graphicx} 
\usepackage{lastpage}
\usepackage{mathtools}
\usepackage{array}
\usepackage{xcolor}
\usepackage{cancel}
\author{Michele Guerrini}
\affiliation
{Carl von Ossietzky Universit\"at Oldenburg, Institute of Physics, 26129 Oldenburg, Germany}
\alsoaffiliation
{Humboldt-Universit\"at zu Berlin, Physics Department and IRIS Adlershof, 12489 Berlin, Germany}

\author{Ana M. Valencia}
\affiliation
{Carl von Ossietzky Universit\"at Oldenburg, Institute of Physics, 26129 Oldenburg, Germany}
\alsoaffiliation
{Humboldt-Universit\"at zu Berlin, Physics Department and IRIS Adlershof, 12489 Berlin, Germany}

\author{Caterina Cocchi}
\affiliation
{Humboldt-Universit\"at zu Berlin, Physics Department and IRIS Adlershof, 12489 Berlin, Germany}
\alsoaffiliation
{Carl von Ossietzky Universit\"at Oldenburg, Institute of Physics, 26129 Oldenburg, Germany}
\email{caterina.cocchi@uni-oldenburg.de}

\title{Long-Range Order Promotes Charge-Transfer Excitations in Donor/Acceptor Co-Crystals}

\begin{document}

%%%%%%%%%%%%%%%%%%%%%%%%%%%%%%%%%%%%%%%%%%%%%%%%%%%%%%%%%%%%%%%%%%%%%
%% The "tocentry" environment can be used to create an entry for the
%% graphical table of contents. It is given here as some journals
%% require that it is printed as part of the abstract page. It will
%% be automatically moved as appropriate.
%%%%%%%%%%%%%%%%%%%%%%%%%%%%%%%%%%%%%%%%%%%%%%%%%%%%%%%%%%%%%%%%%%%%%
\begin{tocentry}
    \centering
    \includegraphics[height=3.5cm]{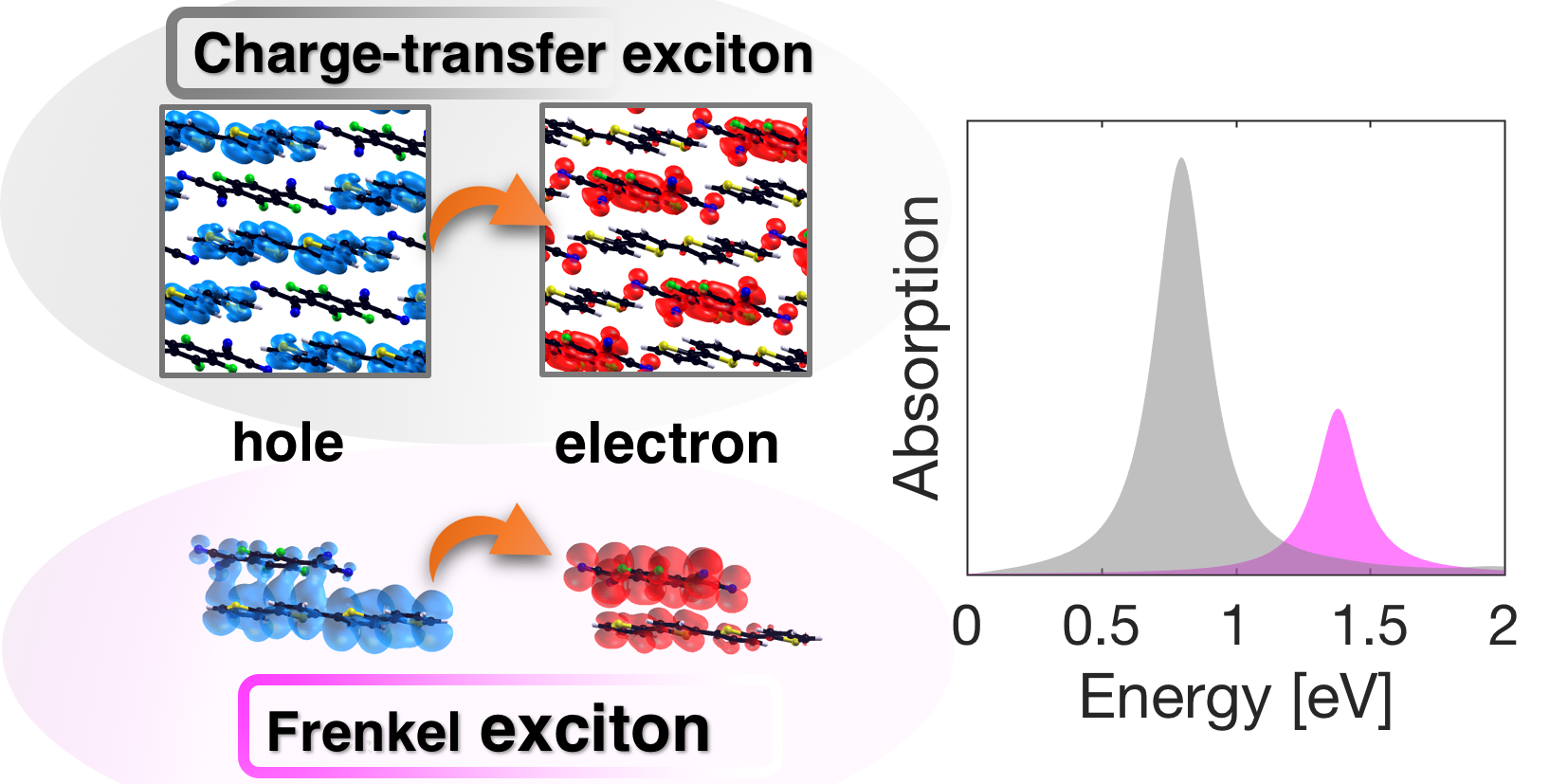}
\end{tocentry}

%%%%%%%%%%%%%%%%%%%%%%%%%%%%%%%%%%%%%%%%%%%%%%%%%%%%%%%%%%%%%%%%%%%%%
%% The abstract environment will automatically gobble the contents
%% if an abstract is not used by the target journal.
%%%%%%%%%%%%%%%%%%%%%%%%%%%%%%%%%%%%%%%%%%%%%%%%%%%%%%%%%%%%%%%%%%%%%
\newpage

\begin{abstract}
Electronic and optical properties of doped organic semiconductors are dominated by local interactions between donor and acceptor molecules.
However, when such systems are in crystalline form, long-range order competes against short-range couplings.
In a first-principles study on three experimentally resolved bulk structures of quaterthiophene doped by (fluorinated) tetracyanoquinodimethane, we demonstrate the crucial role of long-range interactions in donor/acceptor co-crystals.
The band structures of the investigated materials exhibit direct band-gaps decreasing in size with increasing amount of F atoms in the acceptors.
The valence-band maximum and conduction-band minimum are found at the Brillouin zone boundary and the corresponding wave-functions are segregated on donor and acceptor molecules, respectively. 
With the aid of a tight-binding model, we rationalize that the mechanisms responsible for these behaviors, which are ubiquitous in donor/acceptor co-crystals, are driven by long-range interactions.
The optical response of the analyzed co-crystals is highly anisotropic. 
The absorption onset is dominated by an intense resonance corresponding to a charge-transfer excitation.
Long-range interactions are again responsible for this behavior, which enhances the efficiency of the co-crystals for photo-induced charge separation and transport. 
In addition to these results, our study clarifies that cluster models, accounting only for local interactions, cannot capture the relevant impact of long-range order in donor/acceptor co-crystals.
\end{abstract}
\newpage

\newpage
%%%%%%%%%%%%%%%%%%%%%%%%%%%%%%%%%%%%%%%%%%%%%%%%%%%%%%%%%%%%%%%%%%%%%
%% Start the main part of the manuscript here.
%%%%%%%%%%%%%%%%%%%%%%%%%%%%%%%%%%%%%%%%%%%%%%%%%%%%%%%%%%%%%%%%%%%%%
\section{Introduction}

The formation of charge-transfer complexes (CTCs) is one of the key doping mechanisms in organic semiconductors~\cite{goet+14jmcc,salz+16acr,jaco-moul17am}. 
CTCs are characterized by hybridized frontier orbitals between donor and acceptor species and partial charge transfer occurring in the ground state~\cite{zhu+11cm,salz+13prl,mend+15ncom,beye+19cm}. 
Due to this strong electronic coupling, representative systems of this kind, such as \textit{p}-doped oligothiophenes, feature a peculiar type of bright Frenkel excitons that are delocalized over both donor and acceptor moieties~\cite{vale-cocc19jpcc,vale+20pccp}. 
Enhanced charge-transfer character is typically exhibited by higher-energy excitations in the presence of sufficiently long donor chains~\cite{vale-cocc19jpcc}.
This situation is not very favorable to photocurrent generation, in contrast to so-called charge-transfer excitons~\cite{akam-kuro63jcp,zhu+09acr,deib+10am,jail+13natm,macd+20mh}, which feature electrons and holes localized on opposite sides of the donor/acceptor interface~\cite{jail+13natm,liu+16nano}.
However, these excitations are usually weak, due to the small spatial overlap between the involved wave-functions.

The optical properties of CTCs are typically analyzed with the aid of atomistic models based on isolated molecular clusters often including one donor/acceptor pair only, assumed as the key unit of these materials. 
This approximation is well justified in amorphous CTC blends, which are indeed dominated by strong local interactions between donor and acceptor species. The success of corresponding calculations in interpreting experimental data~\cite{mend+13acie,schw+14afm,mend+15ncom,arvi+20jpcb} proves this point.
Nonetheless, cluster models show their limitations when long-range correlations become predominant~\cite{zhu+14jpcc,zhu+15small,hu+16cgd,salz+16cgd,hu+17cec,zhan+18jmcc,sun+19am,macd+20mh}.
These effects are intrinsically present in organic samples with crystalline order, and give rise to the unique electronic and optical characteristics of these materials~\cite{ruin+02prl,buss+02apl,pusc+02prl,tiag+03prb,humm+05pssb,cuda+12prb,cocc-drax15prb,cocc+18pccp}. 

In this paper, we show how long-range interactions impact on the electronic and optical properties of organic donor/acceptor co-crystals.
In this analysis, we consider three structures hosting in their unit cells a quaterthiophene (4T) molecule doped either by one tetracyanoquinodimethane (TCNQ) molecule or by one of its fluorinated derivatives, \ce{F2TCNQ} or \ce{F4TCNQ}.
We perform this study in the \textit{ab initio} framework of density-functional theory (DFT)~\cite{hohe-kohn64pr} and many-body perturbation theory (MBPT)~\cite{onid+02rmp}, with the solution of the Bethe-Salpeter equation (BSE). 
The role of long-range effects prominently emerges already in the band structure of these materials.
The valence-band maximum and the conduction-band minimum appear at the Brillouin zone edge and their corresponding wave-functions are segregated on the donor and acceptor molecules, respectively.
With the aid of a tight-binding model, we explain in a physically transparent manner the fundamental origin of this behavior, and outline general features that are common to donor/acceptor co-crystals. 
The optical response of the analyzed systems is highly anisotropic and, as such, it is described by a full dielectric tensor with non-negligible off-diagonal components.
The optical absorption is dominated by a strong resonance at the onset which corresponds to a charge-transfer excitation: The character of this excited state, stemming directly from the segregated nature of the wave-functions at the frontier, is also determined by long-range correlations.
These findings highlight not only the enhanced electronic and optical response of donor/acceptor co-crystals compared to their amorphous or pseudocrystalline counterparts. They also show the limits of modeling these systems as isolated molecular clusters where the periodic character of the wave-functions is neglected. 

%%%%%%%%%%%%%%%%%%%%%%%%%%%%%%
\section{Methods}
\label{sec:methods}
\subsection{Theoretical Background}
The results presented in this work are obtained from first-principles calculations in the framework of DFT~\cite{hohe-kohn64pr} and MBPT.~\cite{onid+02rmp} 
The key task of DFT consists in solving the Kohn-Sham (KS) equations~\cite{kohn-sham65pr}
\begin{equation}
\label{KSeq}
    \left[-\frac{\nabla^2}{2} + V_{H}[\rho](\textbf{r}) + V_{xc}[\rho](\textbf{r}) + V_{ext}(\textbf{r})\right]\phi^{KS}_{n\textbf{k}}(\textbf{r})=\epsilon^{KS}_{n\textbf{k}} \phi^{KS}_{n\textbf{k}}(\textbf{r}),
\end{equation}
where $V_{ext}(\textbf{r})$ is the external potential from the nuclei, and $V_{H}[\rho](\textbf{r})$ and $V_{xc}[\rho](\textbf{r})$ are the Hartree and the exchange-correlation potentials, respectively. The latter terms are both functionals of the electron density, $\rho(\textbf{r})=\sum_{n\textbf{k}}|\phi^{KS}_{n\textbf{k}}(\textbf{r})|^2$, where $\phi^{KS}_{n\textbf{k}}(\textbf{r})$ are the KS eigenfunctions with eigenenergies $\epsilon^{KS}_{n\textbf{k}}$.

The electronic structure obtained from DFT is the starting point for the solution of the BSE, which is the equation of motion for the electron-hole correlation function~\cite{salp-beth51pr}.
This method enables an accurate description of the optical properties of the materials, including excitonic effects.\cite{stri88rnc} In practice, the solution of the BSE consists in diagonalizing the following eigenvalue problem (the Tamm-Dancoff approximation is assumed):
\begin{equation} \label{bsetda}
    \sum_{jb\textbf{k}'}\mathbb{H}^{BSE}_{ai\textbf{k},bj\textbf{k}'}X^{\xi}_{bj\textbf{k}'}=E_{\xi}X^{\xi}_{ai\textbf{k}},
\end{equation}
where the indexes $i$ and $j$ ($a$ and $b$) run over occupied (unoccupied) states. The effective two-particle BSE Hamiltonian, 
\begin{equation}
    \mathbb{H}^{BSE} = \mathbb{H}^{diag} + \mathbb{H}^{x} + \mathbb{H}^{dir},
\label{eq:H-BSE}
\end{equation}
includes the sum of three terms~\cite{vorw+19es}: The diagonal term, $\mathbb{H}^{diag}$, accounts for the energy difference between single-particle states and, taken alone, it yields the independent particle approximation (IPA); $\mathbb{H}^{x}$ and $\mathbb{H}^{dir}$ include the exchange and direct electron-hole Coulomb integrals, respectively.
The eigenvalues of Eq.~\eqref{bsetda} are the excitation energies. 
The eigenvectors $X_{ai\textbf{k}}^{\xi}$ enter the expression of the transition dipole moment in periodic systems~\cite{Sangalli_jpcm,grosso2000solid}
\begin{equation} \label{tdm}
    \textbf{d}^{\xi}=\sum_{ai} \int_{BZ} X^{\xi}_{ai\textbf{k}} \langle u_a(\textbf{k},\textbf{r})\ket{\hat{i\nabla_{\textbf{k}}}u_i(\textbf{k},\textbf{r})} d\textbf{k},
\end{equation}
where $\Omega$ is the unit-cell volume and $u_{m}(\textbf{k},\textbf{r})$ is the periodic part of the Bloch wave-function with band index $\textit{m}$ and wave-vector $\textbf{k}$.
In turn, Eq.~\eqref{tdm} appears in the expression of the macroscopic dielectric tensor, which in the optical limit ($\mathbf{q} \rightarrow 0$) reads
\begin{equation}
    \varepsilon_{\alpha\beta}(\omega)= \delta_{\alpha\beta} - \frac{4\pi}{\Omega} \sum_{\xi} d_{\alpha}^{\xi}(d_{\beta}^{\xi})^*\frac{2E_{\xi}}{\omega^2 - E_{\xi}^2 + i\omega\eta};
\label{eps_ij}
\end{equation} 
the indexes $\alpha$ and $\beta$ run over the Cartesian coordinates $(x,y,z)$ and $\eta$ is a broadening parameter mimicking the excitation lifetime.

The eigenvectors of Eq.~\eqref{bsetda} additionally provide information about the composition of the excited states~\cite{rohl-loui00prb,cocc-drax15prb}.
As such, they can be used to calculate hole and electron densities defined as
\begin{equation} \label{eq:h}
    \rho_{h}^{\xi} (\textbf{r})=\sum_{ai\textbf{k}} \lvert X_{ai\textbf{k}}\rvert^{2} \lvert \phi_{i\textbf{k}}^{KS}(\textbf{r}) \rvert^{2},
\end{equation}
and
\begin{equation} \label{eq:e}
    \rho_{e}^{\xi} (\textbf{r})=\sum_{ai\textbf{k}} \lvert X_{ai\textbf{k}}^{\xi} \rvert^{2} \rvert \phi_{a\textbf{k}}^{KS}(\textbf{r}) \rvert^{2},
\end{equation}
respectively. 
These definitions are the extensions to periodic systems of those introduced for finite systems~\cite{vale-cocc19jpcc}.
It is worth noting that $\rho_{h}^{\xi}$ and $\rho_{e}^{\xi}$ retain the periodicity of the crystal. 
With the eigenvectors of Eq.~\eqref{bsetda}, we can express also the transition density of the excitation
\begin{equation}
\rho_{TD}^{\xi} (\textbf{r})=\sum_{ai\textbf{k}} X_{ai\textbf{k}}^{\xi} \phi_{a\textbf{k}}^{*KS}(\textbf{r}) \phi_{i\textbf{k}}^{KS}(\textbf{r}),
\label{eq:td}
\end{equation}
which provides information about its polarization and oscillator strength.

%%%%%%%%%%%%%%%%%%%%%%%%%%%%%%%%%%
\subsection{Computational Details}

The donor/acceptor co-crystals investigated in this work correspond to the experimentally resolved geometries reported in Ref.~\citenum{sato+19jmcc}.
DFT calculations on the crystalline structures are performed with the code Quantum Espresso~\cite{gian+17jpcm}. 
A plane-wave basis set with cutoff for wave-functions (electron density) of 50 Ry (200 Ry) and norm conserving pseudopotentials~\cite{Hamann_prb} are used. 
The generalized gradient approximation for the exchange-correlation (xc) functional in the PBE implementation~\cite{perd+96prl} is adopted in conjunction with the Tkatchenko-Scheffler pairwise scheme~\cite{tkat-sche09prl} to account for van der Waals interactions. 
All calculations are performed in the spin-restricted formalism, having checked that the spin polarization plays no role in these systems.
A 4$\times$4$\times$4 \textbf{k}-grid is used to sample the Brillouin zone (BZ) in the solution of the KS equations (Eq.~\ref{KSeq}).
Self-consistent field calculations are performed on top of the optimized geometries obtained by minimizing all interatomic forces until they are smaller than 10$^{-4}$~Ry/\AA{}.

BSE calculations on the co-crystals are subsequently carried out with the code Yambo~\cite{Sangalli_jpcm,marini2009}.
To mimic the quasi-particle correction, the KS electronic structure of each co-crystals is corrected by applying a rigid scissors shift of 1.41~eV (4T-TCNQ), 1.26~eV (4T-F$_2$TCNQ), and 0.96~eV (4T-F$_4$TCNQ) to all conduction bands. 
These values are chosen such that the onsets of the computed spectra are aligned with those measured for corresponding 4T-F$_x$TCNQ blends~\cite{mend+15ncom}. 
The validity of this choice is supported by the overall agreement between our results and those presented in Ref.~\citenum{mend+15ncom} up to 5~eV (see Supporting Information, SI, Figure~S7).
The BSE Hamiltonian (Eq.~\ref{eq:H-BSE}) is formed by a transition space composed of 20 occupied and 40 unoccupied states, and by 4$\times$4$\times$4 \textbf{k}-point grid. 
Its diagonalization is performed using the Haydock-Lanczos algorithm~\cite{Lanczos1950}.
The numerical accuracy on the excitation energies obtained for these calculations is on the order of 10 meV.

The molecular clusters investigated for comparison with the bulk systems are extracted from the experimental crystal structures~\cite{sato+19jmcc} and input into the DFT and MBPT calculations without any further optimization. 
For these tasks, the code MOLGW~\cite{brun+16cpc} is employed, with augmented double-$\zeta$ polarized Gaussian basis set aug-cc-pVDZ\cite{brun12jcp}, frozen-core, and resolution-of-identity~\cite{weig+02} approximations.
The range-separated xc functional CAM-B3LYP~\cite{yana+04cpl} is adopted in these DFT calculations, in order to provide an enhanced starting point to the subsequent MBPT calculations.
In this case, the quasi-particle correction to the KS electronic structure is evaluated explicitly in the single-shot $G_0W_0$ approximation.
Optical spectra and excitations are calculated by solving the BSE in the Tamm-Dancoff approximation. In these calculations, the BSE Hamiltonian is constructed including all the available electron and hole states provided by the adopted basis set.

%%%%%%%%%%%%%%%%%%%%%%%%%%%%
\subsection{Systems}
\label{sec:systems}
\begin{figure}
    \centering
    \includegraphics[width=0.95\textwidth]{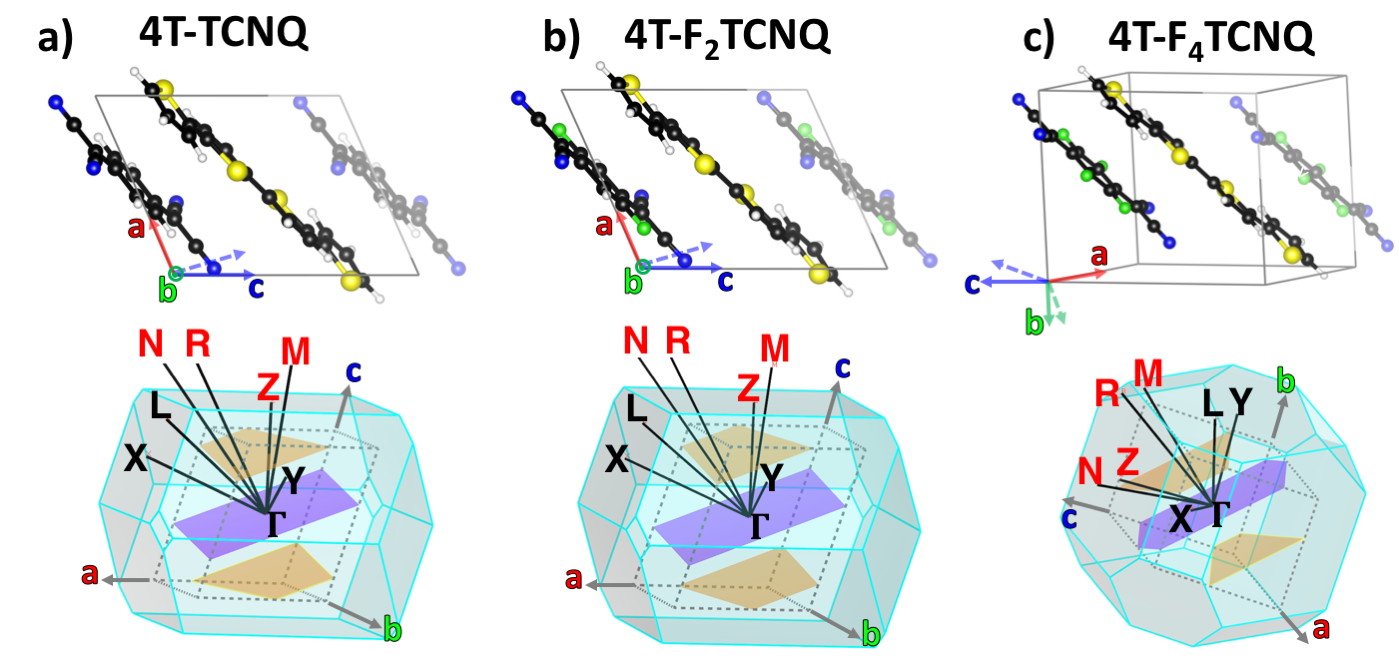}
    \caption{Sketches of the unit cells (upper panels) and Brillouin zones (lower panels) of a) 4T-TCNQ, b) 4T-F$_2$TCNQ, and c) 4T-F$_4$TCNQ co-crystals. In the ball-and-stick representation of the molecules in the unit cells, C, S, N, and H atoms are depicted in black, yellow, blue, and white, respectively. The replica of the acceptor molecule is shaded. In the bottom panels, the unit cells of the co-crystals are indicated by gray dotted lines and the molecular planes of the donor (violet) and acceptor (gold) molecules are highlighted. The black lines represent the vectors connecting $\Gamma$ to the indicated high-symmetry points at the zone edges.}
    \label{fig:systems}
\end{figure}

The three systems considered in this study are the donor/acceptor co-crystals formed by 4T doped by (fluorinated) TCNQ, forming alternated stacks with two molecules per unit cell (see Figure~\ref{fig:systems}, top panel).
Similar to other co-crystals including the same acceptors~\cite{zhu+15small,hu+16cgd,salz+16cgd,hu+17cec}, they are characterized by triclinic structures, which have been recently resolved experimentally~\cite{sato+19jmcc}.
Due to the lack of symmetries, there is no trivial relation between the crystalline axes and the orientation of the molecules in the unit cell. 
As shown in the bottom panel of Figure~\ref{fig:systems}, the stacking direction of the molecules is partially collinear only with the vectors connecting $\Gamma$ to the high-symmetry points $R$, $M$, $Z$, and $N$. 
As discussed in the following, the orientation of the molecules in the unit cell plays a key role in the electronic structure of the co-crystals.

%%%%%%%%%%%%%%
\section{Results and Discussion}
\subsection{Electronic Properties}
\label{sec:electronic}

\begin{figure}
    \centering
    \includegraphics[width=1.0\textwidth]{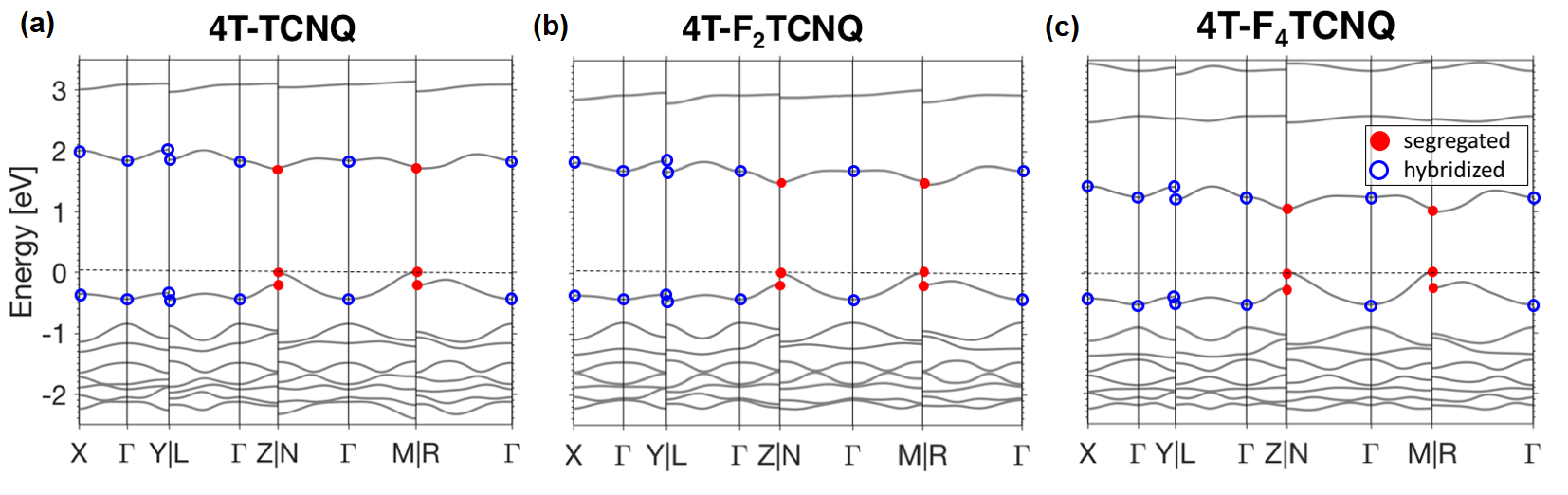}
    \caption{Electronic band structures of (a) 4T-TCNQ , (b) 4T-F$_2$TCNQ and (c) 4T-F$_4$TCNQ co-crystals computed from DFT including a scissors operator to mimic the quasi-particle correction. The valence band maximum is set to zero. Filled and (empty) dots indicate segregated (hybridized) wave-functions at the indicated high-symmetry points.}
    \label{fig:bands}
\end{figure}

The electronic structures computed for the three considered co-crystals are visualized according to the conventional paths proposed in Ref.~\citenum{sety-curt10cms} (see Figure~\ref{fig:bands}).
The band-gap is direct in all systems, and appears at the high-symmetry points $M$ and $N$, where both the highest occupied and the lowest unoccupied states are degenerate. 
The band-gap size decreases monotonically with the amount of fluorination, being 1.74~eV in 4T-TCNQ, 1.49~eV in 4T-F$_2$TCNQ, and 1.04~eV in 4T-F$_4$TCNQ. 
This trend is in line with chemical intuition: F-substitution increases the acceptor strength by increasing its electron affinity~\cite{babu+07cc,kana+09apa,kiva+09cej}.
Except for the variations of the band-gap size, there is no remarkable difference among the band structures of the three co-crystals.
We can exclude that this is an artifact of the underlying PBE functional, since calculations performed with PBE0~\cite{adam-baro99jcp} yield qualitatively the same results (see Figures~S1 and S2 in the SI).

\begin{figure}
    \centering
    \includegraphics[width=0.5\textwidth]{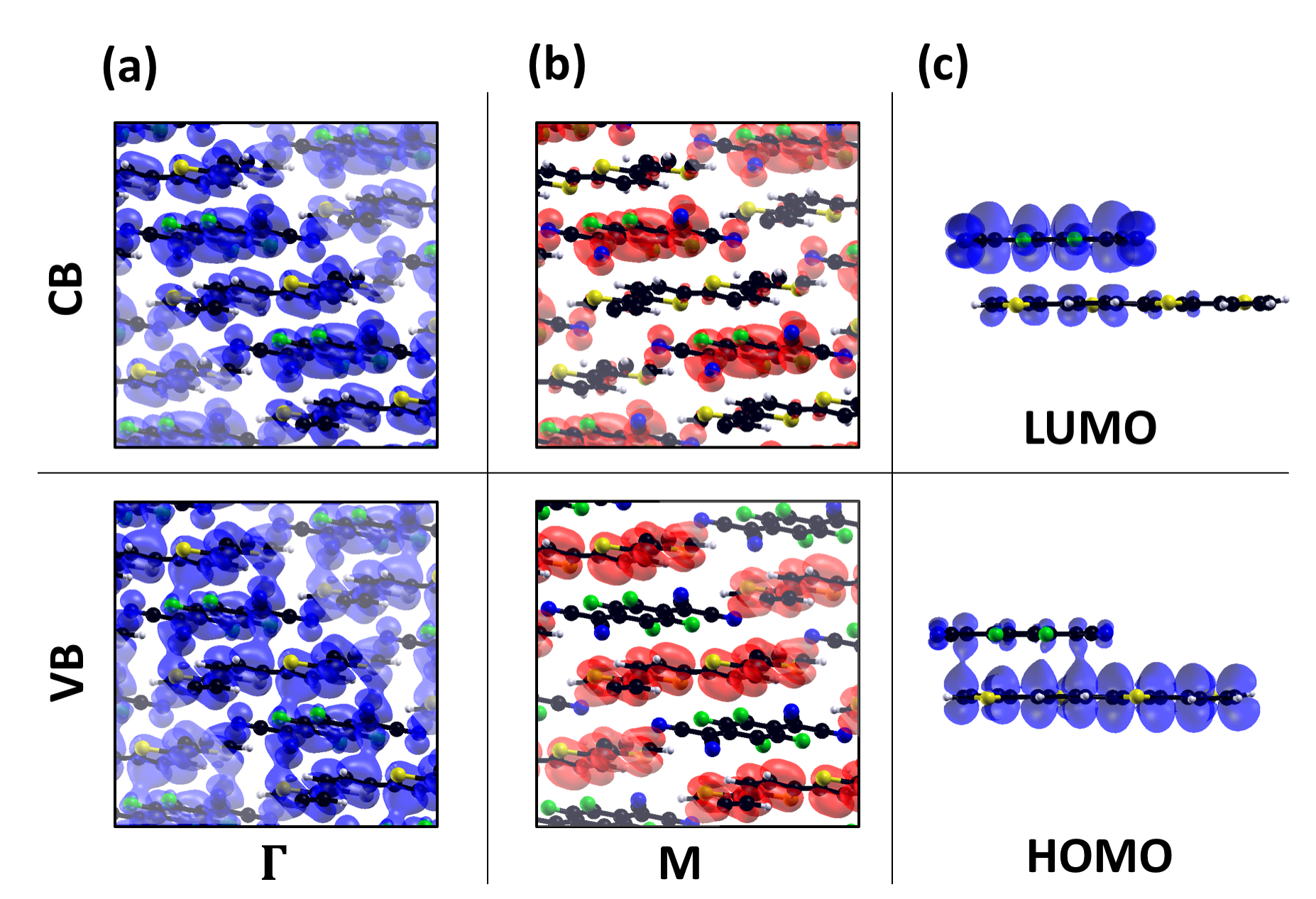}
    \caption{Probability densities of the single-particle wavefunctions of the 4T-F$_4$TCNQ co-crystal at the highest valence (VB) and lowest conduction (CB) bands, evaluated at the high symmetry points (a) $\Gamma$ and (b) $M$. In panel (c) the probability densities of the HOMO and LUMO molecular orbitals of the isolated donor-acceptor bi-molecular complex from the unit-cell of the co-crystal is shown. The isovalue is fixed to 1\% of its maximum value in each plot.}
    \label{fig:wf}
\end{figure}

The character of the wave-functions in the highest valence band (VB) and in the lowest conduction band (CB) varies at different high-symmetry points (see Figure~\ref{fig:bands}). 
\textit{Hybridized states}, marked by hollow circles, appear at $\Gamma$ and at the high-symmetry points $L$, $X$, $Y$, which are connected to the zone center by non-collinear vectors with respect to the stacking direction of the molecules (see Figure~\ref{fig:systems}, bottom panels).
There, the electron density is equally distributed on both molecules in the unit cell [see Figure~\ref{fig:wf}a) for  4T-F$_4$TCNQ and Figures~S5 and S6 in the SI for the other two systems], and the typical bonding (anti-bonding) character of the highest-occupied (lowest-unoccupied) states known for these CTCs~\cite{zhu+11cm,gao+13jmcc,vale-cocc19jpcc,vale+20pccp,mend+15ncom,salz+13prl,krum+21pccp} can be clearly recognized.
This is consistent with the partial charge transfer occurring in the ground state at the interface between donor and acceptor molecules in the co-crystals, which is of the same order, although slightly lower, of that calculated for the isolated complexes in the same geometries as in the crystalline unit cells (see Table~S2 in the SI).

However, at the high-symmetry points $R$, $M$, $Z$, and $N$, which are at the edge of the Brillouin zone (see Figure~\ref{fig:systems}), the VB and CB feature \textit{segregated} wave-functions (red dots in Figure~\ref{fig:bands}), in which the electronic density is localized only on the donor and acceptor molecules, respectively (see Figure~\ref{fig:wf}b).
Notice that also the occupied states in the lower valence-band (VB-1) have the same character as the frontier ones -- see Figures~S4 -- S6 in the SI).
Interestingly, the VB maximum (VBM) and the CB minimum (CBm) correspond to segregated states, meaning that the uppermost occupied state is localized on donor molecules, while the lowest unoccupied state is distributed only on the acceptors.
This scenario is qualitatively very different from the one obtained by modeling the systems with a cluster including only one donor/acceptor pair. 
Even considering the two molecules in the same arrangement as the unit cell of the co-crystal, the highest-occupied and the lowest-unoccupied molecular orbitals correspond to the usual bonding and an anti-bonding states, respectively~\cite{zhu+11cm,gao+13jmcc,vale-cocc19jpcc,vale+20pccp,mend+15ncom,salz+13prl,krum+21pccp} (see Figure~\ref{fig:wf}c and Figures~S4 and S5 in the SI).
As discussed above, this is the character of the VB and CB wave-functions at $\Gamma$ and at other symmetry points, but not throughout the whole Brillouin zone (see Figure~\ref{fig:bands}).
Long-range order in the crystalline structures is responsible for the \textbf{k}-dispersion of the electronic states, and, hence, for their \textbf{k}-dependent character, which leads to \textbf{k}-selective spatial segregation of the wave-functions.
Although counter-intuitive at first sight, neither of these findings is surprising. 
Even in previous studies of CTCs simulated with periodic arrangements, direct band-gaps were found away from $\Gamma$, at the high-symmetry points at the BZ boundary connected to the zone center by vectors aligned with the stacking direction of the donor/acceptor molecules~\cite{beye+19cm,vale+20pccp}.

The ubiquity of these characteristics demands a theoretically robust and yet physically transparent interpretation based on general arguments. 
For this purpose, we introduce a tight-binding (TB) regression model to describe analytically the wave-functions for the VB and CB of the co-crystals on the basis of the hybridized frontier orbitals [HOMO (H) with bonding character and LUMO (L) with anti-bonding character, see Figure~\ref{fig:wf}c)] calculated for the isolated donor/acceptor complex in the respective unit cell.
This choice is motivated by the strong local interactions that couple donor and acceptor molecules in these systems. 
In this way, it is straightforward to connect our results for the co-crystals with those obtained adopting a bimolecular cluster model.

\begin{figure}
    \centering
    \includegraphics[width=0.7\textwidth]{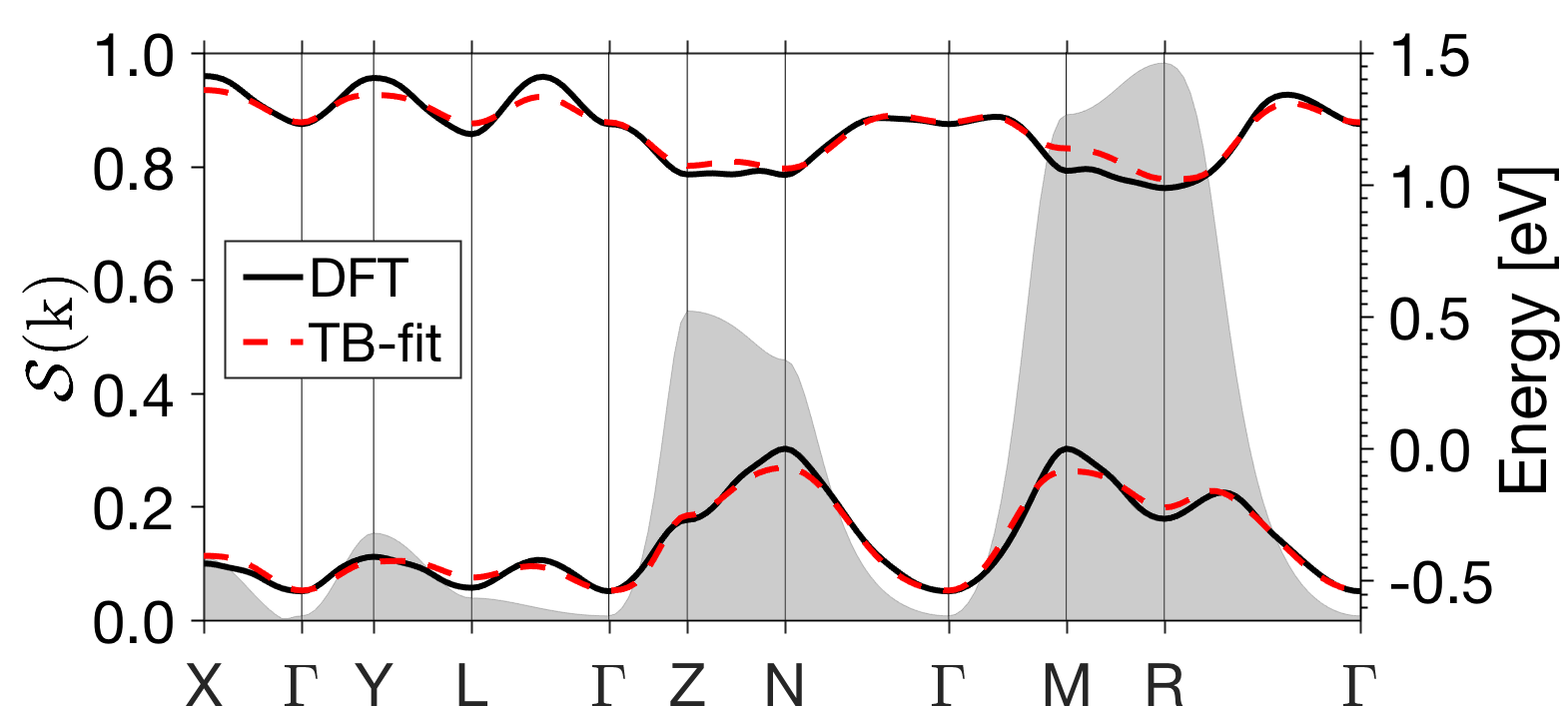}
    \caption{TB fit of the valence (VB) and conduction band (CB) of the 4T-F$_4$TCNQ co-crystal (red dashed lines) compared with the  DFT result (black solid lines). The gray area indicates the \textit{segregation factor}, $\mathcal{S}(\mathbf{k})$.}
    \label{fig:tb}
\end{figure}

In this TB formalism, the wave-functions of the VB and CB in the co-crystals are expressed as~\cite{grosso2000solid}
\begin{equation}
    \label{tb-bloch}
    \psi_{\lambda\textbf{k}}(\textbf{r}) = \sum_j \sum_{m=H,L} C^{\lambda}_m(\textbf{k})\phi_m(\textbf{r} + \textbf{R}_j) e^{-i\textbf{k}\cdot \textbf{R}_j},
\end{equation}
where $\phi_m(\textbf{r})$ are the frontier orbitals of the isolated donor/acceptor bimolecular complex, $C^{\lambda}_m(\textbf{k})$ the orthonormal expansion coefficients for each band $\lambda$ (VB or CB), and $\textbf{R}_j$ the lattice vectors. 
Evidently, $\psi_{\lambda\textbf{k}}(\textbf{r})$ has the periodicity of the lattice and it fulfills Bloch's theorem.
It is worth stressing the \textbf{k}-point dependence of the expansion coefficients which brings about the \textbf{k}-dependent character of the electronic states within each band.
Expressing the Bloch wave-functions in the TB local basis (Eq.~\ref{tb-bloch}) and neglecting the direct overlap integrals between molecular orbitals in different unit cells, band energies and electronic eigenfunctions can be obtained by solving 
\begin{equation}
\begin{pmatrix} A_{HH}(\textbf{k}) & A_{HL}(\textbf{k}) \\ A_{HL}(\textbf{k}) & A_{LL}(\textbf{k})  \end{pmatrix}
\begin{pmatrix} C_H^{\lambda}(\textbf{k}) \\ C_L^{\lambda}(\textbf{k})  \end{pmatrix}
=E_{\lambda}(\textbf{k})
\begin{pmatrix} C_H^{\lambda}(\textbf{k}) \\ C_L^{\lambda}(\textbf{k}) \end{pmatrix},
\label{tbeq}
\end{equation}
where, assuming the nearest-neighbor (NN) approximation, the matrix elements of the TB Hamiltonian, $\mathbb{H}^{TB}$, are calculated as
\begin{equation} \label{Anm}
    A_{nm}(\textbf{k}) = A_{mn}(\textbf{k}) = \sum_{j \in NN} e^{-i\textbf{k} \cdot \textbf{R}_j}\int \phi_n(\textbf{r}) \mathbb{H}^{TB} \phi_m(\textbf{r} + \textbf{R}_j) d^3r = t^{(0)}_{nm} + 2\sum_{j \in NN} t_{nm}^{(j)} \cos(\textbf{k} \cdot \textbf{R}_j),
\end{equation} 
where $t_{nm}^{(j)} = \int\phi_n(\textbf{r}) \mathbb{H}^{TB} \phi_m(\textbf{r} \pm \textbf{R}_j) d^3r = \int\phi_n(\textbf{r} \pm \textbf{R}_j) \mathbb{H}^{TB} \phi_m(\textbf{r}) d^3r$, for $n,m = \{H,L\}$, are the usual on-site ($\textbf{R}_j=0$) and hopping ($\textbf{R}_j = \textbf{R}_a, \textbf{R}_b, \textbf{R}_c$) integrals. 
The bands obtained from the solution of Eq.~\eqref{tbeq} and plotted on top of the DFT result (see Figure~\ref{fig:tb} for 4T-F$_4$TCNQ, and Figure~S13 in the SI for the other two systems) confirm the accuracy of the TB fit and its validity in the subsequent steps. 

From the solution of Eq.~\eqref{tbeq}, the wave-functions of valence and conduction bands result
\begin{equation} \label{eq:vbm}
    \psi_{VB,\textbf{k}}(\textbf{r}) = \sum_j e^{-i\textbf{k} \cdot \textbf{R}_j}W(\textbf{k}) \left[ \phi_H(\textbf{r} + \textbf{R}_j) - \mu(\textbf{k})\phi_L(\textbf{r} + \textbf{R}_j) \right],
\end{equation}
and
\begin{equation} \label{eq:cbm}
    \psi_{CB,\textbf{k}}(\textbf{r}) = \sum_j e^{-i\textbf{k} \cdot \textbf{R}_j}W(\textbf{k}) \left[ \mu(\textbf{k})\phi_H(\textbf{r} + \textbf{R}_j) + \phi_L(\textbf{r} + \textbf{R}_j) \right],
\end{equation}
respectively, while the energy dispersion relations in the valence and conduction bands are
\begin{equation} \label{eq:Ek}
    E_{\lambda}(\textbf{k})= \sum_{j }\sum_{n,m} C^{\lambda}_n(\textbf{k}) C^{\lambda}_m(\textbf{k}) t^{(j)}_{nm} \cos(\textbf{k} \cdot \textbf{R}_j)   \;\;\;\; [\lambda = VB, CB].
\end{equation}
In Eqs.~\eqref{eq:vbm} and \eqref{eq:cbm}, we introduced (see SI for further details)
\begin{equation}
     \mu(\textbf{k}) = \frac{A_{HL}(\textbf{k})}{E_{(-)}(\textbf{k}) + \sqrt{E^2_{(-)}(\textbf{k}) + A^2_{HL}(\textbf{k})}}, 
     \label{eq:mu}
\end{equation}
with $E_{(-)}(\textbf{k}) = A_{LL}(\textbf{k}) - A_{HH}(\textbf{k}) > 0$ and
\begin{equation}
W(\textbf{k}) = \frac{1}{\sqrt{1 + \mu^2(\textbf{k})}}.
     \label{eq:w}
\end{equation}
The expansion coefficients in Eq.~\eqref{tb-bloch} can then be expressed as:
\begin{equation}
 C^{VB}_H(\textbf{k})=\frac{1}{\sqrt{1 + \mu^2(\textbf{k})}},\;\;\;\;\; C^{VB}_L(\textbf{k})= -\frac{\mu(\textbf{k})}{\sqrt{1 + \mu^2(\textbf{k})}}  , 
 \label{eq:C-VB}
\end{equation}
and  
\begin{equation}
 C^{CB}_H(\textbf{k})=\frac{\mu(\textbf{k})}{\sqrt{1 + \mu^2(\textbf{k})}},\;\;\;\;\; C^{CB}_L(\textbf{k})=\frac{1}{\sqrt{1 + \mu^2(\textbf{k})}}.  
 \label{eq:C-CB}
\end{equation}
With the aid of Eqs.~\eqref{eq:mu} and \eqref{eq:w}, we introduce the \textit{segregation factor}
\begin{equation}
    \mathcal{S}(\mathbf{k})= 1 - W^2(\textbf{k})\left( 1 - |\mu(\textbf{k})| \right)^2 = 1 - \frac{\left( 1 - |\mu(\textbf{k})|\right)^2}{1 + \mu^2(\textbf{k})},
    \label{eq:sf}
\end{equation}
which evaluates how much of $\psi_{VB,\textbf{k}}(\textbf{r})$ and $\psi_{CB,\textbf{k}}(\textbf{r})$ is localized on the donor and acceptor molecules, respectively.
By definition, $\mathcal{S}(\mathbf{k}) \in [0,1]$. 

From $\mathcal{S}(\mathbf{k})$ computed for the 4T-F$_4$TCNQ co-crystal (see Figure~\ref{fig:tb}), we notice the following behavior which is common also to the other two systems (see Figure~S13 in the SI): In the vicinity of the high-symmetry points $M$, $R$ as well as $Z$ and $N$, $\mathcal{S}(\mathbf{k})$ has values between 0.5 and 1, which are indicative of segregation, whereas $\mathcal{S}(\mathbf{k})\rightarrow 0$ close to $X$, $\Gamma$, $Y$, and $L$, where the wave-functions are equally delocalized between donors and acceptors.
We can rationalize these results going back to the equations introduced above.
From Eq.~\eqref{eq:mu}, it is apparent that $\mu(\mathbf{k})$ is maximized when $A_{HL}(\textbf{k})$ is large. This happens at the high-symmetry points $M$, $R$, $Z$, and $N$, which are along \textbf{k}-vectors that are collinear with $\textbf{R}_c$, \textit{i.e.}, the unit cell vector aligned with the stacking direction of the molecules (see Figure~\ref{fig:systems} and Table~S6 in the SI).
In this case, the off-diagonal elements of the TB Hamiltonian (Eq.~\ref{tbeq}) are non-negligible and, consequently, the expansion coefficients of the Bloch wave-function, $C^{\lambda}_m(\textbf{k})$ -- see Eq.~\eqref{tb-bloch} -- are similar in magnitude.
In this scenario, long-range interactions, stemming from the periodic nature of the co-crystals, overcome local interactions between donor and acceptor molecules and ``mix'' together the hybridized HOMO and LUMO of the donor/acceptor pair  (see Figure~\ref{fig:wf}c). 
This interference is destructive and, as such, drastically reduces the delocalization of the electronic states enhancing instead their localization on the donor (acceptor) molecules.
On the other hand, at the high-symmetry points $X$, $Y$, and $L$, which are along \textbf{k}-vectors approximately perpendicular to the stacking direction of the molecules in the unit cell (see Figure~\ref{fig:systems} and Table~S6 in the SI), $A_{HL}(\textbf{k})$ tends to vanish and the HOMO-LUMO mixing is negligible.
In the opposite limiting case of $A_{HL}(\textbf{k}) =0$, the TB Hamiltonian (Eq.~\ref{tbeq}) is diagonal, and one coefficient $C^{\lambda}_m(\textbf{k})$ is always dominant.
Consequently, $\mu(\mathbf{k}) \rightarrow 0$ implies that $W(\textbf{k}) \rightarrow 1$ (see Eq.~\ref{eq:w}), and eventually $\mathcal{S}(\mathbf{k}) \rightarrow 0$: Local interactions dominate over the long-range couplings, and the bonding (anti-bonding) character of the HOMO (LUMO) in the donor/acceptor pair is preserved in the VB and CB wave-functions of the co-crystals.

Another point that can be rationalized with the aid of the TB model is why the VBM and the CBm in the co-crystals are not at $\Gamma$. 
The size of the band-gap and its position in reciprocal space are determined by the on-site and hopping integrals $t_{nm}^{(j)}$ in Eq.\eqref{Anm}. 
Two necessary conditions have to be fulfilled (see SI for the derivation steps): First, the band-gap is at a stationary point in \textbf{k}-space, \textit{i.e.}, $\nabla_{\textbf{k}} \left[ E_{CB}(\textbf{k}) - E_{VB}(\textbf{k}) \right]=0$, which is the standard condition to determine high-symmetry points in reciprocal space; Second, the following relation holds: %
\begin{equation} \label{neccondgap}
    \left[ \frac{t_{LL}^{(j)} - t_{HH}^{(j)}}{2}E_{(-)}(\textbf{k}) + t_{HL}^{(j)}A_{HL}(\textbf{k}) \right]\cos(\textbf{k} \cdot \textbf{R}_j) < 0  \;\;\;\;\;  \textrm{for}  \;\;\;\;\; j=a,b,c.
\end{equation}
The on-site and hopping integrals obtained for the considered co-crystals (see Table~S7 in the SI) indicate that this condition is indeed fulfilled at the high-symmetry points $M$ and $N$.
On the other hand, it is never satisfied at $\Gamma$, as the term on the left-hand-side of Eq.~\eqref{neccondgap} is always positive for $\mathbf{k}=0$.
Note that, while Eq.~\eqref{neccondgap} is general, the region in reciprocal space where it is fulfilled depends on the details of the system and, in particular, on the primitive lattice vectors, and on the on-site and hopping integrals. 
We also recall that the presented model stands upon the initial assumption that the crystalline wave-functions can be expanded onto the local basis of the HOMO and LUMO of the donor/acceptor pair in the unit cell (Eq.~\ref{tb-bloch}). 
Under these premises, the TB model and the arguments developed in the analysis herein are applicable also to other donor/acceptor co-crystals.

%%%%%%%%%%%%%%%%%%%%%%%%%%%%%%%%%%%%%%%%
\subsection{Optical Properties}
\label{sec:optics}

\begin{figure*}
    \centering
    \includegraphics[width=1.0\textwidth]{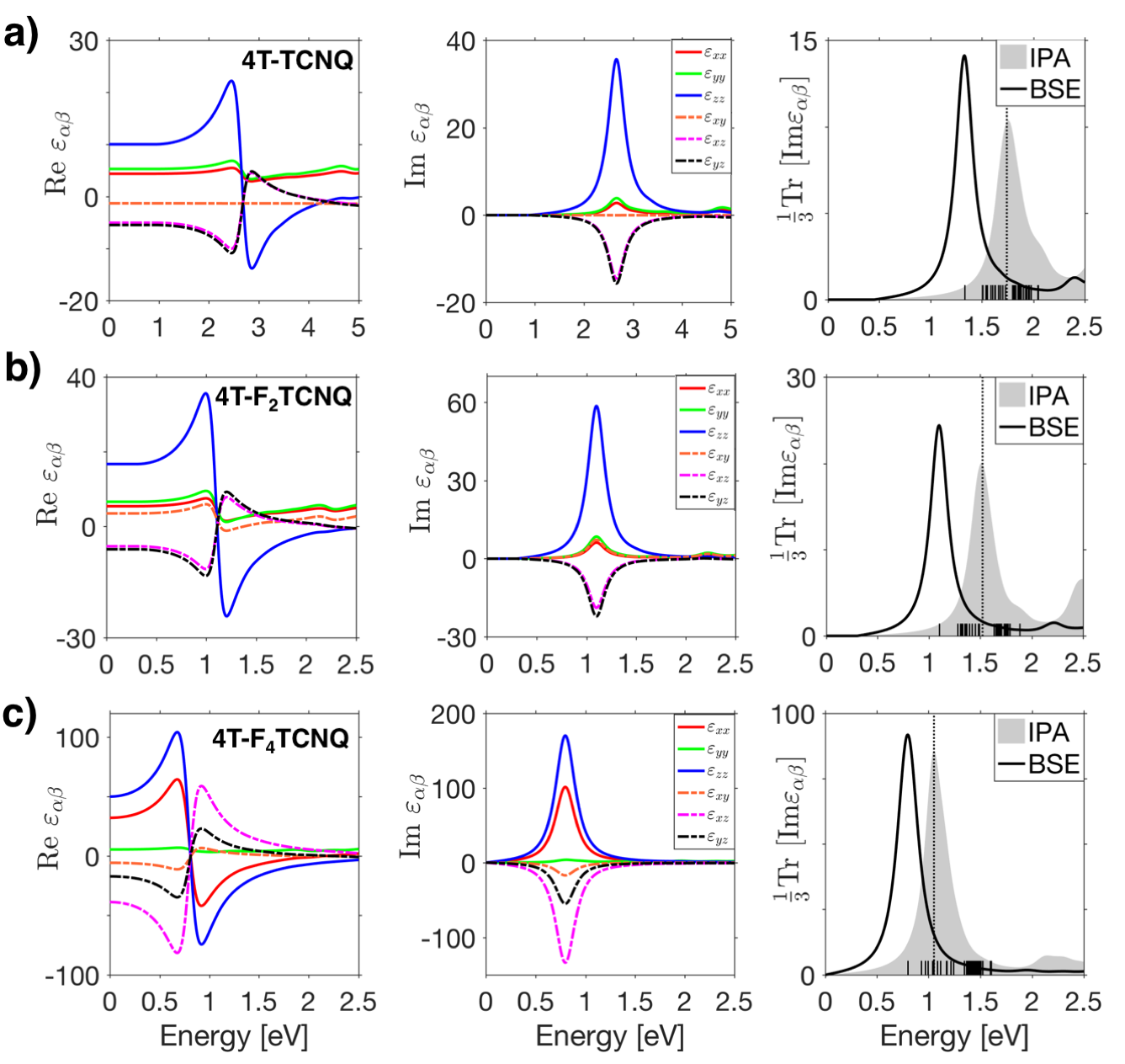}
    \caption{Real (left) and imaginary (center) parts of the dielectric tensor calculated with the BSE for a) 4T-TCNQ, b) 4T-F$_2$TCNQ and c) 4T-F$_4$TCNQ co-crystals; diagonal (off-diagonal) components are indicated by solid (dashed) lines. On the right panels the average absorption spectra computed from the trace of the diagonal components are shown, as obtained from the BSE (black line) and from the independent particle absorption (IPA, gray area). The black bars mark the discrete spectrum of BSE solutions and the vertical dashed lines the band-gap energy coinciding with the lowest-energy excitation in the IPA. A Lorentzian broadening of 200 meV is applied to all BSE and IPA spectra.}
    \label{fig:opt}
\end{figure*}

With the knowledge gained on the electronic structure, we continue our analysis with the optical properties of the investigated co-crystals.
The anisotropy of the triclinic lattices is reflected in the optical response of these materials, which is expressed by six inequivalent components of the macroscopic dielectric tensor (see Figure~\ref{fig:opt}, left and middle panels).
In all systems, the $\varepsilon_{zz}$ component is the dominant one, in agreement with the fact that the $z$-axis is approximately aligned with the stacking direction of the molecules in the unit cell (see Figure~\ref{fig:systems}). 
In the 4T-F$_4$TCNQ co-crystal, where the organic constituents are oriented slightly differently than in the other two systems, also $\varepsilon_{xx}$ has a non-negligible magnitude.
On the other hand, $\varepsilon_{yy}$ is small in all spectra.
$\textrm{Re}\,\varepsilon(\omega=0)$ represents the static dielectric permittivity which increases with the amount of fluorination in the acceptor (Figure~\ref{fig:opt}, left panels). 
This is consistent with the fact that F-substitution increases the number of electrons in the systems, thereby enhancing the electronic screening.
Off-diagonal components of $\varepsilon_{\alpha\beta}(\omega)$ are, in general, non-negligible. 
In the imaginary part, they give rise to negative peaks that contribute to the absorption under specific experimental conditions concerning the mutual orientation of the sample and the incident beam and/or the light polarization.~\cite{cocc+18pccp}. 
Such conditions can be reproduced by diagonalizing the calculated tensor in the
appropriate reference frame~\cite{yeh1980ss,pusc-ambr06aenm,vorw+16cpc,pass-paar17josab}. 

In the absence of a polarization-resolved experimental point of comparison for the computed spectra of the 4T-TCNQ, 4T-F$_2$TCNQ, and 4T-F$_4$TCNQ co-crystals, in Figure~\ref{fig:opt}, right panels, we visualize the absorption by plotting the trace of the diagonal components of the imaginary part of $\varepsilon_{\alpha\beta}(\omega)$ depicted in the middle panels. 
All spectra exhibit at the onset a distinct resonance in the infrared.
The increase of fluorination in the acceptor leads also to a red-shift of this peak from approximately 1.3~eV in 4T-TCNQ to 0.8~eV in 4T-F$_4$TCNQ (see Table S3 in the SI), with a concomitant enhancement of the oscillator strength (see Figure~\ref{fig:opt}). 
This behavior is in agreement with measurements on crystallites with the same composition~\cite{mend+15ncom}. 
Also higher-energy peaks, including the one assigned to a localized excitation on 4T, are qualitatively featured by our results (see Figure~S7 in the SI).
On the other hand, polaronic features that are present in the measured spectra~\cite{mend+15ncom,arvi+20jpcb} cannot be captured with the adopted BSE formalism, where only vertical transitions (\textit{i.e.}, transitions in which the atomic positions optimized in the ground state are held fixed also in the excited states) can be computed.

The BSE eigenvalues obtained from Eq.~\eqref{bsetda} are plotted as vertical bars in the spectra in Figure~\ref{fig:opt}, right panels, in order to outline the energy distribution of the excited states.
In this way, it is clear that the first peak corresponds to the lowest-energy excitation, which is bright in all systems (see SI, Table~S3).
Higher-energy excitations encompassed by the first peak have low oscillator strength and, therefore, do not significantly contribute to the absorption.
For comparison, we plot for each system also the spectra computed from the IPA where the exchange and the direct electron-hole Coulomb interactions are neglected. 
The vertical dotted bars plotted on the right panels of Figure~\ref{fig:opt} correspond to the direct band gaps of the co-crystals (see Figure~\ref{fig:bands}) and coincide with the lowest-energy solution of the diagonal part of the BSE Hamiltonian (Eq.~\ref{eq:H-BSE}) which yields the IPA.
In this way, we can estimate the exciton binding energy of the first bright excitation as the difference between its energy obtained by solving the BSE and in the IPA.
We find that the exciton binding energy decreases with increasing number of F atoms in the acceptor: in 4T-TCNQ and 4T-F$_2$TCNQ, it amounts to 420~meV, while in 4T-F$_4$TCNQ it decreases to 250~meV. 
This result is consistent with the enhanced electronic screening when an increasing amount of electrons are added to the system.
Notably, the sharp peak dominating the BSE spectra appears also in the IPA results with similar strength, as commonly occurring in organic semiconductors~\cite{ruin+02prl,humm-ambr05prb,cuda+12prb,cocc-drax15prb} regardless of the presence of ground-state charge transfer.
This indicates that this resonance is intrinsic of the electronic structure of the materials and that, contrary to conventional inorganic semiconductors or insulators~\cite{lask+05prb,arna+06prl}, it is not built up by the screened electron-hole attraction.
Excitonic effects act on the considered co-crystals mainly by lowering the energy of the first absorption peak and by slightly enhancing its oscillator strength.

\begin{figure}
    \centering
    \includegraphics[width=0.5\textwidth]{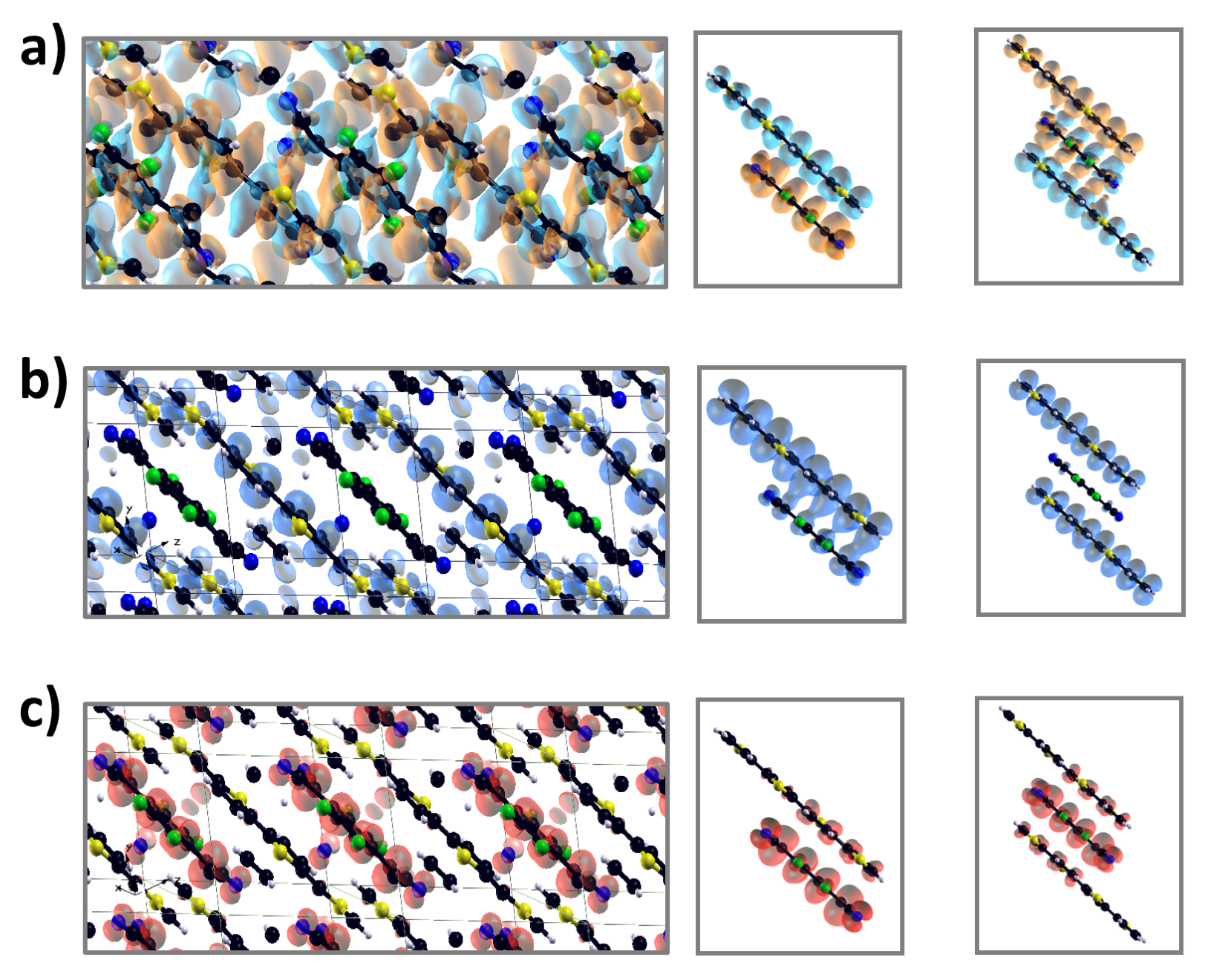}
    \caption{a) Transition density, b) hole density, and c) electron density computed for the first excited state of the 4T-F$_4$TCNQ co-crystal (left) as well as for the isolatd bi- and trimolecular donor/acceptor clusters extracted from it (middle and right panel). In panel a) cyan and orange isosurfaces indicate positive and negative charge-density domains, respectively.}
    \label{fig:eh}
\end{figure}

To gain further insight into the role of long-range order on the optical absorption of the investigated co-crystals, we analyze the character of the first bright excitation in 4T-F$_4$TCNQ, as representative of all systems (corresponding results for the 4T-TCNQ and 4T-F$_2$TCNQ co-crystals are shown in Figure~S9 in SI). 
The computed transition density (Eq.~\ref{eq:td}) is displayed in Figure~\ref{fig:eh}a), left panel: The alternation of positive and negative charge-density domains along the stacking direction of the molecules indicates a correspondingly aligned transition-dipole moment.
Considering now for comparison the transition density calculated for the isolated donor/acceptor pair in the unit cells, we notice an analogous orientation of the transition-dipole moment with electron depletion on the donor and electron accumulation on the acceptor [see Figure~\ref{fig:opt}a), middle panel].
The picture does not change even when we consider the trimolecular cluster containing an alternation of donor/acceptor/donor molecules arranged as in the crystal unit cell.
These results are in line with those obtained in Ref.~\citenum{vale+20pccp} for ideally aligned non-periodic 4T-F$_4$TCNQ and 6T-F$_4$TCNQ CTCs, and indicate that local interactions between donor and acceptor molecules are responsible for the polarization of the first bright excitation.
This correct prediction of the cluster models is one of the main reasons for their success in reproducing the optical spectra of CTCs.
Clearly, in the co-crystal, the relative oscillator strength of the excitation is larger than in the molecular clusters (compare Figure 5 and Figure S12 in the SI), due to the enhanced extension and
overlap between the periodic wave-functions therein.

To better understand the peculiarity of the lowest-energy excitation in the absorption spectra of the co-crystals, which has large oscillator strength and concomitantly charge-transfer character, we evoke once again the TB model introduced earlier to rationalize the behavior of the electronic states.
To this end, we evaluate the matrix elements in the transition dipole moment introduced in Eq.~\eqref{tdm} using Eq~\eqref{tb-bloch}, \textit{i.e.}, in the basis of the frontier orbitals of the donor/acceptor pair.
In this way, we obtain
\begin{equation} \label{tdm2}
    \langle u_a(\textbf{k},\textbf{r})\ket{\hat{i\nabla_{\textbf{k}}}u_j(\textbf{k},\textbf{r})} = \sum_{m=H,L} C_m^a(\textbf{k}) \; (i\nabla_{\textbf{k}}) \; C_m^j ({\textbf{k}}) + \sum_{m,m'=H,L} C_{m'}^a({\textbf{k}})C_m^j({\textbf{k}}) \textbf{d}_{m'm} ,
\end{equation}
where 
\begin{equation} \label{tdm3}
    \textbf{d}_{m'm} = \int \phi_{m'} (\textbf{r}) \; \textbf{r} \; \phi_m (\textbf{r}) \; d^3r.
\end{equation}
Both terms on the right hand side of Eq.~\eqref{tdm2} contain the expansion coefficients introduced in Eq.~\eqref{tb-bloch} and made explicit in Eqs.~\eqref{eq:C-VB} and \eqref{eq:C-CB}.
However, while the second term is dominated by the dipole moment calculated on the basis of the frontier orbitals of the bimolecular cluster (see Eq.~\ref{tdm3}), the first one contains the variation of the expansion coefficients with respect to \textbf{k} and, as such, enables long-range coupling between the periodic electronic states even when their spatial overlap is small.
Thanks to this contribution, even charge-transfer excitations can have large oscillator strength in the co-crystals.
We emphasize that this could not happen in isolated molecular clusters which feature non-periodic and, thus, \textbf{k}-independent wave-functions.

Inspecting now the hole and electron densities (see Eqs.~\ref{eq:h} and \ref{eq:e}, respectively) calculated for the first excitation in the co-crystals, we notice remarkable differences with respect to the corresponding non-periodic CTCs. 
As shown in Figure~\ref{fig:eh}b)-c) for 4T-F$_4$TCNQ and in Figure~S9 for the other two materials, in the co-crystal, these quantities are segregated on the donor and acceptor molecules only, similarly to the wave-functions at the VBM and CBm [see Figure~\ref{fig:wf}a)-b)], while in the molecular clusters they reflect the delocalized bonding and anti-bonding character of the frontier orbitals [see Figure~\ref{fig:wf}c) and Figure~S10 in the SI].  
These findings suggest that long-range order and long-range interactions, as they cause the segregation of the single-particle states at the frontier, also determine the localization of hole and electron densities on donor and acceptor molecules, respectively.
Evidently, these effects can be captured only by simulating the co-crystals explicitly accounting for the periodicity of their wave-functions.
Any non-periodic representation, regardless of how many molecules it includes, will never fulfill the purpose.

%%%%%%%%%%%%%%%%%%%%%%%%%
\section{Conclusions}
\label{sec:conclu}

In summary, we have investigated from first-principles many-body theory the electronic and optical properties of three recently synthesized donor/acceptor co-crystals formed by quarterthiophene doped by TCNQ, \ce{F2TCNQ}, and \ce{F4TCNQ}.
All the examined systems feature a direct band gap that decreases in size with increasing F-content in the acceptor molecules.
The VBM and CBm are located away from $\Gamma$, and the corresponding wave-functions are localized on donor and acceptor molecules, respectively. 
With the aid of a tight-binding regression model, we have rationalized these results: The electronic wave-functions are segregated on the donor/acceptor molecules at high-symmetry points that are along \textbf{k}-vectors largely aligned with the stacking direction of the molecules.
In this case, long-range interactions prevail over local couplings, and unmake the bonding and anti-bonding character of the HOMO and the LUMO in the donor/acceptor pairs.
Long-range order impacts also the optical properties.
The structural anisotropy of the lattice is reflected in the photoresponse of the systems, which is characterized by a dielectric tensor with six independent, non-vanishing components.
The absorption spectra of the co-crystals are characterized by an intense resonance at the onset polarized along the stacking direction of the molecules, the energy of which decreases with increasing F-content in the acceptor.
The associated exciton binding energy decreases with the amount of fluorination in the acceptor, ranging from 420~meV in 4T-TCNQ and 4T-\ce{F2TCNQ} to 250~meV in 4T-\ce{F4TCNQ}, consistently with the increased electronic screening that is also responsible for the decrease of the band-gap and the increase of the static permittivity.
The hole and electron densities of the first, bright excitation reflect the segregated character of the wave-functions at the VBM and CBm, thereby highlighting the charge-transfer nature of this excited state. 

The results of this work have important implications regarding the first-principles description of donor/acceptor co-crystals. 
While cluster models remain a valid and computationally efficient tool to capture local interactions~\cite{vale+20pccp} and thereby to qualitatively access the optical absorption of CTCs, neglecting the \textbf{k}-dependence of the electronic bands prevents reproducing the character of the frontier states and hence the nature of the first bright excitation.
The inclusion of long-range interactions by accounting for the periodicity of the co-crystals is therefore essential to capture their electronic and optical properties on a quantitative level. 

Our findings are relevant also from a broader perspective.
The intense, charge-transfer excitation at lowest-energy is the distinct signature of the examined co-crystals.
Differently from amorphous donor/acceptor blends where the absorption associated to charge-transfer excitations is typically very weak, due to the negligible orbital overlap between the constituting species~\cite{jail+13natm,liu+16nano}, the delocalized character of the electronic wave-function in the co-crystals ensures the coexistence of large oscillator strength and electron-hole separation.
While it is established that crystallinity improves the \textit{charge transport} of the samples~\cite{veho+10jacs,ortm+11pssb}, our results indicate that long-range order enhances also the \textit{charge transfer} character of the lowest-energy excitation.
Increasing the quality of the co-crystals is therefore expected to enhance the ability of these materials to produce photo-generated charge current.

%%%%%%%%%%%%%%%%%%%%%%%%%%%%%%%%%%%%%%%%%%%%%%%%%%%%%%%%%%%%%%%%%%%%%
%% The "Acknowledgement" section can be given in all manuscript
%% classes.  This should be given within the "acknowledgement"
%% environment, which will make the correct section or running title.
%%%%%%%%%%%%%%%%%%%%%%%%%%%%%%%%%%%%%%%%%%%%%%%%%%%%%%%%%%%%%%%%%%%%%
\begin{acknowledgement}
This work was funded by the German Research Foundation (DFG) under project FoMEDOS, Project Numbers 182087777 (CRC 951) and 286798544 – HE 5866/2-1, by the German Federal Ministry of Education and Research (Professorinnenprogramm III) and by the State of Lower Saxony (Professorinnen für Niedersachsen). M.G. acknowledges additional financial support by Fondazione Delle Riccia. Computational resources were provided by the North-German Supercomputing Alliance (HLRN), project bep00076.
\end{acknowledgement}

%%%%%%%%%%%%%%%%%%%%%%%%%%%%%%%%%%%%%%%%%%%%%%%%%%%%%%%%%%%%%%%%%%%%%
%% The same is true for Supporting Information, which should use the
%% suppinfo environment.
%%%%%%%%%%%%%%%%%%%%%%%%%%%%%%%%%%%%%%%%%%%%%%%%%%%%%%%%%%%%%%%%%%%%%
\begin{suppinfo}
In the Supporting Information we report additional information about the structural, electronic, and optical properties of the co-crystals, as well as of the bi- and trimolecular clusters extracted from them. We also provide further details and the fitting parameters of the tight-binding model.

\end{suppinfo}

%%%%%%%%%%%%%%%%%%%%%%%%%%%%%%%%%%%%%%%%%%%%%%%%%%%%%%%%%%%%%%%%%%%%%
%% The appropriate \bibliography command should be placed here.
%% Notice that the class file automatically sets \bibliographystyle
%% and also names the section correctly.
%%%%%%%%%%%%%%%%%%%%%%%%%%%%%%%%%%%%%%%%%%%%%%%%%%%%%%%%%%%%%%%%%%%%%
%\bibliography{bib}

\providecommand{\latin}[1]{#1}
\makeatletter
\providecommand{\doi}
  {\begingroup\let\do\@makeother\dospecials
  \catcode`\{=1 \catcode`\}=2 \doi@aux}
\providecommand{\doi@aux}[1]{\endgroup\texttt{#1}}
\makeatother
\providecommand*\mcitethebibliography{\thebibliography}
\csname @ifundefined\endcsname{endmcitethebibliography}
  {\let\endmcitethebibliography\endthebibliography}{}

\end{document}